\documentclass[journal=jctcce,manuscript=article,layout=traditional]{achemso}
\usepackage[T1]{fontenc}
\usepackage[version=3]{mhchem}
\usepackage[dvipsnames]{xcolor}
\usepackage{graphicx}
\usepackage{subcaption}
\usepackage{amssymb}
\usepackage{overpic}
\usepackage{soul}
\usepackage{bm}
\usepackage{enumerate}
\usepackage{physics}
\usepackage{hyperref}
\usepackage{subcaption}

\hypersetup{pdfborderstyle={/S/U/W 0.5}}

\title[]{
Charge Transfer with a Spin. II: A Framework for Diabatization which Localizes Charge and Spin}
\author{Alok Kumar}
\affiliation{Department of Chemistry, University of Pennsylvania, Philadelphia, Pennsylvania, 19104, USA}
\alsoaffiliation{Department of Chemistry, Princeton University, Princeton, New Jersey, 08540, USA}
\author{Zhen Tao}
\affiliation{Department of Chemistry, University of Rhode Island, Kingston, Rhode Island, 02881,
USA}
\author{Zuxin Jin}
\affiliation{Department of Chemistry, University of Pennsylvania, Philadelphia, Pennsylvania, 19104, USA}
\author{Joseph E. Subotnik}
\affiliation{Department of Chemistry, Princeton University, Princeton, New Jersey, 08540, USA}
\author{Tian Qiu}
\email{tianq@princeton.edu}
\affiliation{Department of Chemistry, Princeton University, Princeton, New Jersey, 08540, USA}

\begin{document}
 
\maketitle
\begin{abstract}
    We investigate a diabatization procedure that localizes charges (in real space) and localizes spins  (in spin space) for open-shell systems that exhibit charge transfer  in the presence of spin-orbit coupling. The procedure is applied to a two-state crossing between pairs of Kramers-restricted doublet states (which can also be considered effectively a four-state crossing).  
    To generate the relevant diabatic states, we employ a two-step optimization over complex-unitary rotations that sequentially maximizes dipole and spin moments through iterative Jacobi sweeps; the resulting update rules are effectively equivalent to those of approximate joint diagonalization (AJD) applied to charge and spin. The method converges rapidly and yields smooth diabatic potential energy surfaces that preserve dipole and spin properties (e.g., a smoothly varying spin quantization axis) along the reaction coordinate while maintaining time-reversal symmetry. 
\end{abstract}

\section{Introduction}
Charge‐transfer (CT) processes play a central role in a wide range of phenomena from photosynthetic electron transfer\cite{wasielewski1992photoinduced, rochaix2011regulation, gust2001mimicking} and molecular photovoltaics\cite{coropceanu2019charge, ostroverkhova2016organic} to spin–orbit–coupled phenomena in transition-metal complexes\cite{may2024new, zobel2021quest} and chiral molecular systems \cite{naaman2019chiral, eckvahl2023direct, naaman2024can}. In many of these systems, multiple electronic states interact strongly with nuclear motion \cite{newton1983electron, marcus:sutin:1985, worner2017charge}, and in open-shell systems, additional coupling between spin and orbital angular momentum introduces nontrivial spin dynamics. These couplings make the adiabatic representation, though convenient for electronic structure calculations, ill-suited for describing charge localization and nonadiabatic population transfer \cite{marchi1993diabatic, hsu1997sequential}.
Accurate modeling of electron transfer processes often demands the construction of diabatic states, which simplify the description of state-to-state couplings and charge localization.

Constructing diabatic states that remain physically meaningful in the presence of spin–orbit coupling (SOC) presents two distinct challenges, as SOC breaks spin conservation and entangles spatial and spin degrees of freedom. First, traditional diabatization schemes, including Boys/GMH localization\cite{boys:1960, fosterboys:1960,  cave:1996:gmh, subotnik:2008:boysgmh}, Edmiston–Ruedenberg\cite{edmiston1963localized}, fragment charge diabatization and fragment energy diabatization\cite{voityuk:2002:fcd, voityuk:2014:ftdm, hsu:2004:sf,hsu:2006:tt, hsu2008characterization}, or block-diagonalization \cite{venghaus2016block, wittenbrink2016new} approaches, are designed for spin-free adiabatic states and therefore cannot handle the degenerate (or nearly degenerate) complex-valued spinors required to describe open-shell systems with strong SOC. 
Interestingly, fragment spin difference \cite{you2010fragment}(which is designed for spin-processes) 
does not distinguish between different spin states of a given spin manifold. 
Previous studies of multistate density functional theory (MSDFT) have mainly focused on spin-multiplet energy splittings \cite{grofe2017spin, grofe2017diabatic, liu2025multistate}, with limited treatment of how the spin character evolves with molecular geometry.
Second, although SOC is commonly introduced as a fixed perturbation on spin-free electronic states, in reality its magnitude and phase evolve with nuclear geometry, reflecting the continuous mixing of spin and orbital angular momenta \cite{domcke2006relativistic, marian2012spin, thorning2022geometry}, necessitating extreme attention to phases and degeneracies (with many crossings possible). One previous approach to  diabatization in the presence of SOC (by Zeng \cite{zeng2017diabatization}) sought to construct and follow smooth SOC matrix elements along nuclear distortions by following the character of the one-electron orbitals, as pioneered by Atchity and Ruedenberg \cite{rued:1997:diabat}) and  Truhlar and co-workers\cite{truhlar:2001:fourfold, truhlar:2014:hoyer_boys_quad}, and rotating to a given reference set of states.  
This approach is viable, but will be tedious for large molecules, especially if we include SOC before diagonalization (and thus must be much more careful when tracking diabatic states).

With this background in mind, our goal in this manuscript is to develop a diabatization procotol as appropriate for charge transfer processes where there are non-innocent spin degrees of freedom. In particular, if we seek to generalize Marcus theory to handle spin-dependent electron transfer, note that the most important starting point for any fermi golden rule calculation is the assignment of an initial state. Indeed, among the many possible sets of diabatic states, for our purposes, we would like to find that diabatic state that is most stable and most likely to serve as an initial or final spin-dependent electronic state.\cite{subotnik:2009:erdiabat}
Of course, localizing spin (in spin space) means choosing diabatic states whose spin-quantization axis varies smoothly along the reaction coordinate, rather than twisting arbitrarily with geometry. A slowly and smoothly varying spin axis ensures that changes in the spin expectation values reflect genuine spin–orbit–driven physics, rather than arbitrary choices of gauge, especially within a degenerate manifold. In other words, if we can calculate such a choice of spin-dependent diabatic state, we should have a reasonably clear view of the electronic spin-state during nuclear motion (provided we know where the electron is).

With this intuition in mind, below we present a unified diabatization framework that  localizes both charge and spin through unitary rotations, all in the  presence of SOC. Our
procedure builds upon Boys localization conceptually. While the Boys approach was originally formulated for orbitals\cite{fosterboys:1960, edmiston1963localized} and later extended to many-electron diabatic states\cite{subotnik:2008:boysgmh}, we make the jump here to systems with spin.  Physically speaking, standard Boys localization localizes charges and can be derived by assuming that we are interested in the states of a quantum system surrounded and mixed together fully by a random electric dipole in an unknown direction arising from an  electrostatic environment\cite{subotnik:2009:erdiabat},
\begin{equation}
    \hat{H}=-{\bf \mu} \cdot \mathcal{E}.
\end{equation}
Similarly, the proposed algorithm below localized spin can be derived in a similar manner by assuming a random magnetic dipole in an unknown direction arising from a magnetostatic environment
\begin{equation}
    \hat{H}=-{\bf S} \cdot \mathcal{B}.
\end{equation}

In what follows, we will apply both standard and spin Boys localization to 
a set of adiabatic states for charge transfer in presence of SOC. 
We note that this (second) paper
follows an initial (first) paper\cite{kumar2026chargetransferspini} where the relevant adiabatic states were constructed. 
The final result of our analysis will be a set of charge and spin localized diabatic states with smooth energies and spin texture.  

An outline of this article is as follows. In Sec. \ref{sec:method}, 
we review the adiabatic-to-diabatic basis transformation and motivate why molecules with an odd number of electrons are an interesting system to study spin diabatization. In Sec. \ref{sec:diabatization_cs},  we present our two-step, Boys-localization inspired, diabatization procedure in the context of a four  state problem, 
with two spatial states and two spin, corresponding to a spin-dependent electron transfer. We also 
present numerical results plotting the diabatic energy surfaces and spin expectation values in  Sec. \ref{sec:results}.  In Sec. \ref{sec:discussion}, we discuss the subtle but significant difference between spin symmetry and time-reversal symmetry and what that difference means in context of spin expectation values. Finally, in Sec. \ref{sec:conclusion}, we conclude and highlight the future dynamical implications of a slowly, smoothly non-adiabatic smooth diabatic basis as far as understanding radical electron transfer.

\section{Method}\label{sec:method}

\subsection{Adiabatic and diabatic basis}
Within the Born--Oppenheimer approximation, the full molecular Hamiltonian can be written as
\begin{align} \label{eq:Hbo}
    \hat{H}(\bm{r},\bm{R}) = \hat{T}_{\text{nuc}} + \hat{H}_{\text{el}}(\bm{r};\bm{R}),
\end{align}
where $\bm{r}$ and $\bm{R}$ denote electronic and nuclear coordinates, respectively, $\hat{T}_{\text{nuc}}$ is the nuclear kinetic energy operator, and
$\hat{H}_{\text{el}}(\bm{r};\bm{R})$
is the electronic Hamiltonian for clamped nuclei. For each fixed nuclear geometry, the adiabatic electronic states $\ket{\psi_k(\bm{R})}$ are defined as the eigenfunctions of the electronic Hamiltonian, 
\begin{align}
    \hat H_{\mathrm{el}}\,|\psi_k\rangle
= E_k\,|\psi_k\rangle,
\end{align}
and the associated eigenvalues $E_k$ define the adiabatic potential energy surface.
These adiabatic states form an orthonormal basis at each geometry and are the
natural output from any electronic-structure calculation. However, as is well-known\cite{tvv:2010:arpc, shu2022diabatic, truhlar:2014:hoyer_boys_quad}, these adiabatic states usually do not coincide with the chemically intuitive electronic configurations that one would like to follow along a reaction coordinate. Moreover,  in the presence of external degrees of freedom like nuclear vibrations or solvent environment, these adiabatic eigenstates of the isolated system may not be stationary states\cite{subotnik:2009:erdiabat}. 
Lastly, in the vicinity of avoided crossings, 
the charge distribution (and thus the dipole moment expectation values) as well as the spin expectation values can vary sharply. For all of these reasons, and in order to model electron transfer \`a la Marcus theory\cite{marcus:1956}, there is a strong reason to consider a different, diabatic electronic basis, which has largely constant electronic character and is stabilized by the environment.

To obtain states that are better aligned with chemical intuition, the standard approach today\cite{shu2022diabatic, subotnik:2015:acr} is to exploit the unitary freedom within a chosen adiabatic manifold $\{\ket{\psi_k}\}$  (of $n$ states) and construct
a diabatic basis $\{\ket{\xi_i} \}$ via a geometry-dependent unitary
transformation,
\begin{align}\label{eq:diab_state}
	\ket{\xi_i} = \sum_{j}^{n} \ket{\psi_j} U_{ji}
\end{align}
where $\bm{U}$ is a unitary matrix of size $n \times n$. Above, $\bm{U}$ is usually called an adiabatic-to-diabatic transformation.

\subsection{Diabatization For Electron Transfer (Without Spin): Boys localization}
Let us temporarily ignore electronic spin. When electron transfer happens, we typically need two diabatic states to represent the initial and final states of electron transfer \cite{tvv:2010:arpc}. The Boys localization scheme was originally developed for localizing orbitals \cite{boys:1960, rued:1963} and was later extended to the localization of many-electron states \cite{cave:1996:gmh,subotnik:2008:boysgmh}. We begin with a set of adiabatic states $\{\ket{\psi_k}\}$ and define a unitary rotation matrix $\bm{U}$ as in Eq. \ref{eq:diab_state}
such that it maximizes the magnitude of dipole moment difference between diabatic states. 
\begin{align}\label{eq:func_boys_diff}
    f_{\rm Boys}(\bm{U}) =    \sum_{
    i,j
    }  \left |\bra{\xi_i}\hat{\vec{\mu}}\ket{\xi_i} - \bra{\xi_{j}}\hat{\vec{\mu}}\ket{\xi_{j}} \right |^2
\end{align}
where $\hat{\mu}$ is the electric dipole operator.

The maximization in Eq. \ref{eq:func_boys_diff} is  equivalent to maximizing the sum of the magnitudes of the dipole moments of all the states
\begin{align}\label{eq:func_boys_diag}
    f_{\rm Boys}(\bm{U}) \equiv \sum_{i} \left |\bra{\xi_i}\hat{\vec{\mu}}\ket{\xi_i}\right|^2
\end{align}

As evident from Eq.~\ref{eq:func_boys_diag}, the Boys objective function can be written as a sum of the squares of the diagonals of  a collection of dipole matrices ($\bm{\mu}_x, \bm{\mu}_y, \bm{\mu}_z $) in a diabatic basis; maximizing such an objective function is therefore equivalent to seeking a unitary rotation  that (approximately) diagonalizes all three matrices. An equivalent viewpoint is to cumulatively minimize the off-diagonal elements, leading to the approximate joint diagonalization (AJD) formulation for a set of Hermitian matrices. Closed-form AJD updates have been developed for complex unitary rotations and successfully used for orbital localization in presence of SOC\cite{cardoso1996jacobi,ciupka2011localization}.

In practice, in order to maximize the function in Eq. \ref{eq:func_boys_diag}, one usually  follows the approach in Ref. \citenum{edmiston1963localized}.  Namely,  instead of optimizing the full unitary matrix at once, we perform successive 2$\times$2 rotations within subspaces spanned by pairs of adiabatic states, for which the Boys function can be maximized exactly.  Iterating this pairwise optimization over all state pairs and repeating the sweep until changes in the objective are negligible yields a unitary transformation that gives {\em approximately} diagonal dipole moment matrices.   In practice, the procedure above, with so-called “Jacobi sweeps’’, usually converge fairly quickly reaching a maximum after only a few sweeps, regardless of the number of charge centers or states \cite{subotnik:2008:boysgmh}. 
The resulting diabatic states usually have small (though certainly not zero) derivative couplings.\cite{alguire:2013:closs_for_yarkony} 

\subsection{Odd-electron molecule and Kramers degeneracy}
The requirements of a diabatization scheme become more involved in the presence of spin because of Kramers' degeneracy. For a system with an odd number of electrons, a theorem due to Kramers, states that in the absence of external magnetic fields, every energy level is at least doubly degenerate \cite{kramers1930theorie}. These degeneracies arise from the invariance of the Hamiltonian under the anti-linear time-reversal symmetry (TRS) operator $\hat{T}$. If the Hamiltonian $\hat{H}$ is time-reversal symmetric,
\begin{align}\label{eq:H_trs}
    [\hat{H},\hat{T}] = 0,
\end{align}
and $\ket{\psi}$ is an eigenstate with energy $E$, then $\hat{T}\ket{\psi}$ is also an eigenstate with the same energy,
\begin{align}
    \hat{H}\ket{\psi} = E \; \ket{\psi}
    \quad \Rightarrow \quad
    \hat{H}\hat{T}\ket{\psi} = E \; \hat{T}\ket{\psi}.
\end{align}

The two degenerate states are related by the time-reversal operator which satisfies $\hat{T}^2 = -1$.
Such a pair, $\{\ket{\psi}, \ket{\bar{\psi}}\}$, forms a \emph{Kramers pair} and obeys
\begin{align}
    \hat{T} \ket{\psi} &= \ket{\bar{\psi}}, \\
    \hat{T^2} \ket{\psi} &= -\ket{\psi}, \\
    \braket{\psi}{\bar{\psi}} &= 0.
\end{align}

The states in a Kramers pair span a two-dimensional degenerate subspace. Within this subspace, there is no unique choice of orthonormal basis: any pair of orthonormal linear combinations is equally valid and yields the same energies and other time-reversal-even observables (dipole, spin-orbit coupling etc.). Physically, this internal freedom can be viewed as the freedom to attach a different spin-quantization axis to the doublet. By applying geometry-dependent unitary rotations that mix the two members of a Kramers pair, one  leaves the overall spectrum unchanged while continuously changing the direction of the local spin expectation value.
If this freedom is not controlled, the spin along a reaction coordinate is masked by such arbitrary gauge rotations, obscuring a clear interpretation of how spin can couple to
nuclear motion. 

\begin{figure}[ht]
    \centering
    \includegraphics[width=1.0\textwidth]{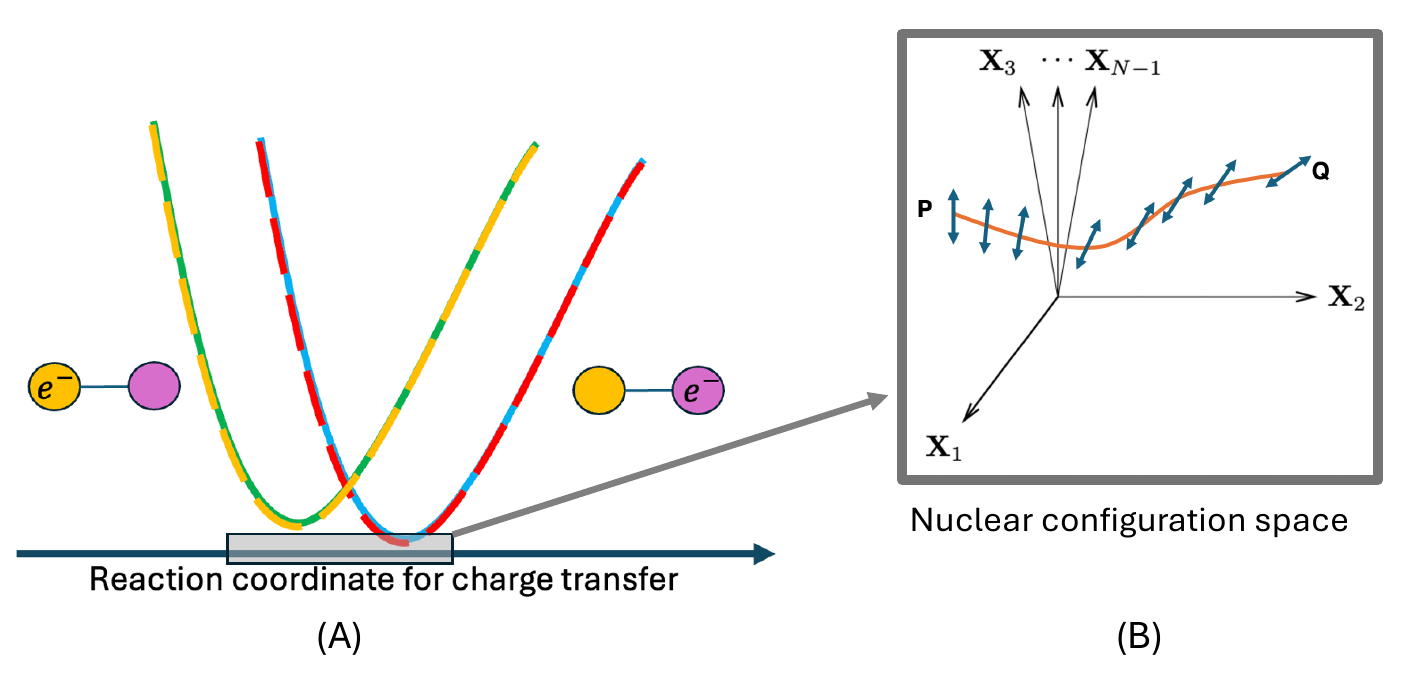}
    \caption{Schematic illustration of charge transfer and the spin-quantization axis in an odd-electron system. (A) Doubly degenerate diabatic potential energy surfaces corresponding to charge being localized on left (or right fragment) as shown on yellow (magenta) balls. (B) Multi-dimensional nuclear configuration space. A nuclear trajectory from configuration P to Q (orange curve) defines an effective reaction coordinate; along this path, the spin-quantization axis of a Kramers doublet (blue arrows) can rotate.}
    \label{fig:CTspin_nuc}
\end{figure}

Vice versa, by enforcing a slowly and smoothly varying spin-quantization axis—for instance, via a spin-localization criterion that minimizes the off-diagonal components of the spin operator within each Kramers pair—we effectively establish a gauge convention. In this gauge, the local spin expectation value for each diabat  varies slowly and smoothly with nuclear geometry and can be interpreted as a well-defined geometric observable. 
Fig. \ref{fig:CTspin_nuc} (Panel A) shows a schematic of a charge transfer process where a molecule (and presumably its environment)  undergoes a nuclear rearrangement (vibrations, rotations, etc.) as it goes through curve crossing. Imagine a molecule with $N$ atoms, with 3$N$ total degrees of freedom, and  3$N$-3 internal degrees of freedom (after removing the center-of mass; see Panel B, where each axis is 3-dimensional).  During a rearrangement, in the presence of SOC, the  spin-quantization axis of Kramers pair can rotate as a function of molecular geometry. 

\section{Diabatization of charge and spin}\label{sec:diabatization_cs}
Let us now imagine we are diabatizing a system with 2N electronic states, arranged in Kramers' pairs  as described above.  Our goal is to construct quasi-diabatic states for charge-transfer systems with SOC that simultaneously satisfy three goals: (i) we must localize the charge, (ii) we must localize the spin,  and (iii) we must preserve time-reversal symmetry. Because of the nature of the algorithm discussed above, our proposed diabatization can be defined by focusing one at a time on a given pair of states (to be rotated together). 

To that end, consider two adiabatic states, 
$\ket{\psi_1}$ and $\ket{\psi_2}$ respectively.  These are orthonormal eigenstates of the electronic Hamiltonian, and  their time-reversed partners can be defined as
\begin{align}
    \ket{\bar{\psi}_k} = \hat{T}\ket{\psi_k}, \qquad k \in \{1,2\}.
\end{align}
Each Kramers pair $\{\ket{\psi_k},\ket{\bar{\psi}_k}\}$ forms an orthonormal doublet.
Moreover, if $\{\ket{\psi_1},\ket{\psi_2}\}$ is an orthonormal set, then their respective time-reversed partners $\{\ket{\bar{\psi}_1},\ket{\bar{\psi}_2}\}$ are also orthonormal, since
\begin{align}
    \braket{\bar{\psi}_1}{\bar{\psi}_2}
    = \braket{\hat{T}\psi_1}{\hat{T}\psi_2}
    = (\braket{\psi_1}{\psi_2})^*
\end{align}
Thus the four adiabatic states $\{\ket{\psi_1},\ket{\psi_2},\ket{\bar{\psi}_1},\ket{\bar{\psi}_2}\}$ can be chosen as an orthonormal basis of the relevant 4-dimensional subspace,
and the relevant diabatic basis $\{\ket{\xi_1},\ket{\xi_2},\ket{\bar{\xi}_1},\ket{\bar{\xi}_2}\}$ is obtained from the adiabatic basis by a unitary transformation. 
Since the adiabatic eigenstates are two-component spinor wavefunctions with complex-valued coefficients (due to the presence of spin–orbit coupling), the adiabatic–to–diabatic transformation matrix $\bm{U}$ must be taken as a general complex-valued unitary matrix rather than a purely real orthogonal rotation as in Eq. \ref{eq:func_boys_diff}.
By restricting $\bm{U}$ to the subgroup that preserves the Kramers structure (i.e., acting with SU(2) rotations within each Kramers pair and mixing the two doublets in a TRS-consistent way), we ensure that the diabatic states remain orthonormal and organized into Kramers pairs, while achieving simultaneous charge and spin localization along the reaction coordinate. See section \ref{sec:jacobi_sweep} for more details about preserving time-reversal symmetry during  adiabatic-to-diabatic transformation.

\subsection{Optimization functions} \label{sec:opt_func}
We employ a two-step diabatization scheme to achieve simultaneous charge and spin localization. In the first step, we maximize the dipole expectation values, followed by the maximization of the spin expectation values. 
The two functions being maximized are: 
\begin{equation} \label{eq:opt_dipole}
\textbf{Localize dipole:} \qquad 
 f_1(\bm{U}) = \sum_i \sum_{\alpha = x,y,z} 
\left| \langle \xi_i \mid \hat{\mu}_\alpha \mid \xi_i \rangle \right|^2 
+ \left| \langle \bar{\xi_{i}} \mid \hat{\mu}_\alpha \mid \bar{\xi_{i}} \rangle \right|^2
\end{equation}

\begin{equation}
\label{eq:opt_spin}
\textbf{Localize spin:} \qquad 
f_2^i(\bm{U}) = \sum_{\alpha = x,y,z} 
\left| \langle \xi_i \mid \hat{S}_\alpha \mid \xi_i \rangle \right|^2 
+ \left| \langle \bar{\xi_{i}} \mid \hat{S}_\alpha \mid \bar{\xi_{i}} \rangle \right|^2
\end{equation}
where $\bm{\hat{\mu}}_\alpha$ and $\bm{\hat{S}}_\alpha$ are the Cartesian dipole moment and spin matrices, respectively. Note that the optimization for localizing dipole is performed over all four states simultaneously (Eq. \ref{eq:opt_dipole}) whereas spin localization (for each charge diabatized state \textit{i}) is between time reversal pairs only (Eq. \ref{eq:opt_spin}).
We follow an iterative procedure of Jacobi sweeps to find a sequence of optimal rotation matrix $\bm{U}$ 
that maximizes these functions while preserving time-reversal symmetry.

\subsection{Jacobi sweeps with Time Reversal Symmetry} 
In the absence of external magnetic fields or other time–reversal–breaking
interactions, the electronic Hamiltonian commutes with the time–reversal operator $\hat{T}$ (Eq. \ref{eq:H_trs}).
Our diabatization procedure should
not artificially break this symmetry. This requirement makes each Jacobi rotation
slightly more constrained than in a generic complex Jacobi sweep.

\label{sec:jacobi_sweep}
The rotation matrix $\bm{U}$ is a complex unitary matrix that transforms the
adiabatic basis to the diabatic one.  We construct $\bm{U}$ as a product of
sequential Jacobi rotations, each acting in a low–dimensional subspace spanned
by a small number of adiabatic states,
\begin{align}
  \bm{U} = \prod_m \bm{Q}_m ,
  \label{eq:Uprod}
\end{align}
where each $\bm{Q}_m$ differs from the identity only within the selected subspace. At this point, it is useful to distinguish two types of rotations:

\paragraph{(i) Rotations within a single Kramers pair: Spin diabatization.}

Consider a Kramers doublet $(i,\bar{i})$ belonging to a given charge–localized
state (where $\ket{\bar{i}}  = \hat{T}\ket{i}$). We label the states using just the indices for brevity.
In the ordered basis $(i,\bar{i})$, the corresponding $2\times 2$ block of
$\bm{Q}_m$ is
\begin{align}
  \bm{Q}_m(\gamma,\phi)
  =
  \begin{bmatrix}
    \cos\gamma      & -\sin\gamma\,e^{-i\phi} \\
    \sin\gamma\,e^{i\phi} & \cos\gamma
  \end{bmatrix},
  \label{eq:Q_spin}
\end{align}
with all other matrix elements equal to those of the identity.  Because
$\ket{\bar{i}} = \hat{T}\ket{i}$ and $\hat{T}$ is antiunitary, one can easily show that any such
$\mathrm{SU}(2)$ rotation automatically preserves the Kramers relation
between the two rotated states, i.e. $\ket{\bar{i}'}  = \hat{T}\ket{i'}$.  In our calculations we perform one such
rotation in each charge–localized state.
To avoid any confusion with the ``spin-diabat'' terminology as used in previous literature \cite{granucci2012surface, mai2015general, mukherjee2021modeling}, we emphasize that the states obtained from our spin-diabatization procedure are not fixed-spin states (i.e. they are not eigenfunctions of $S^2$). Rather, they are spin-localized states whose local spin-quantization axis, and hence spinor wavefunctions, vary (smoothly) with nuclear geometry.

\paragraph{(ii) Rotations between two different Kramers pairs: Dipole
diabatization.}
Dipole localization mixes two distinct Kramers pairs, say $(i,\bar{i})$ and
$(j,\bar{j})$.  Here time–reversal symmetry requires that we rotate the four
states $\{\ket{i},\ket{j},\ket{\bar{i}},\ket{\bar{j}}\}$ together: an
arbitrary $2\times 2$ rotation between $\ket{i}$ and $\ket{j}$ must be
accompanied by a corresponding rotation between their time–reversal partners
$\ket{\bar{i}}$ and $\ket{\bar{j}}$, otherwise the updated states would no
longer satisfy $\ket{\bar{\xi}_i} = \hat{T}\ket{\xi_i}$ and the exact Kramers
degeneracy would be spuriously broken.

In practice, we work with a $4\times 4$ Jacobi matrix that is block–diagonal
in the ordered basis $(i,j,\bar{i},\bar{j})$,
\begin{align}
  \bm{Q}_m(\gamma,\phi)
  =
  \begin{bmatrix}
    \cos\gamma      & -\sin\gamma\,e^{-i\phi} & 0 & 0 \\
    \sin\gamma\,e^{i\phi} & \cos\gamma        & 0 & 0 \\
    0 & 0 &
    \cos\gamma      & -\sin\gamma\,e^{i\phi} \\
    0 & 0 &
    \sin\gamma\,e^{-i\phi} & \cos\gamma
  \end{bmatrix},
  \label{eq:Q_dipole}
\end{align}
where the upper–left block mixes $(i,j)$ and the lower–right block mixes their
time–reversal partners $(\bar{i},\bar{j})$ with the conjugate phase.  

\subsection{Solving for optimal rotation matrix}
Finally, for both the dipole and spin diabatization steps, we note that the $2\times 2$ or $4\times 4$ rotational steps involve complex-valued matrices  (Eq. \ref{eq:Q_spin} and \ref{eq:Q_dipole}, respectively) for which the solutions in Ref. \citenum{rued:1963} are not immediately applicable. Nevertheless, the solutions
$\gamma$ and $\phi$ for each Jacobi matrix $\bm{Q}_m$ can be obtained by solving a
cubic equation that maximizes the corresponding localization functional. See Appendix (Sec. \ref{sec:si}) for more details.

Below, as an example, we will 
focus on a pair of two Kramers pairs, together with the TRS
constraints above. For such a problem, we require only three (rather than  $4 \times \frac{3}{2} =6$) independent $4\times 4$ rotations
coupling $(i,j)$, $(i,\bar{j})$, and $(j,\bar{i})$ in the dipole. 
Note that, without TRS, for a generic four–state problem, we would require  $n(n-1)/2 = 6$ independent distinct pairwise rotations. 
Thereafter, in line with Sec. \ref{sec:opt_func} above, after we localize the charges,  we calculate 
two independent $2\times 2$ rotations within  the $(i,\bar{i})$ and $(j,\bar{j})$ subspaces
so as to localize the spin.

\section{Numerical 
Results}\label{sec:results}
We consider an open–shell system with an odd number of electrons undergoing charge transfer in presence of SOC.  As an illustrative example, we focus on the phenoxy-phenol (ph-ph) system, in which charge is transferred between two phenyl fragments as the bridging hydrogen atom moves from one fragment to the other (See Fig. \ref{fig:pes_adiab_vs_diab} (A)). The electronic structure  for this CT process can be described in terms of two adiabatic states, which in absence of any external magnetic field, are doubly degenerate by the Kramers theorem. The present study builds on Paper~1 \cite{kumar2026chargetransferspini}, where the two lowest adiabatic doublets were analyzed in detail using Dynamically-weighted State-Averaged Constrained CASSCF (eDSC/hDSC) which is  generalized to include complex-valued orbitals. We have implemented the energy calculations and diabatization algorithm in a development version of Q-Chem \cite{epifanovsky2021software}. 
This phenoxy-phenol(ph-ph) system has been studied before within a restricted open-shell framework \cite{qiu2024fast, qiu2024efficient} and the adiabatic states are the same as obtained in Paper 1.

\subsection{Diabatic states}
\begin{figure}[ht]
    \centering
    \includegraphics[width=1.0\textwidth]{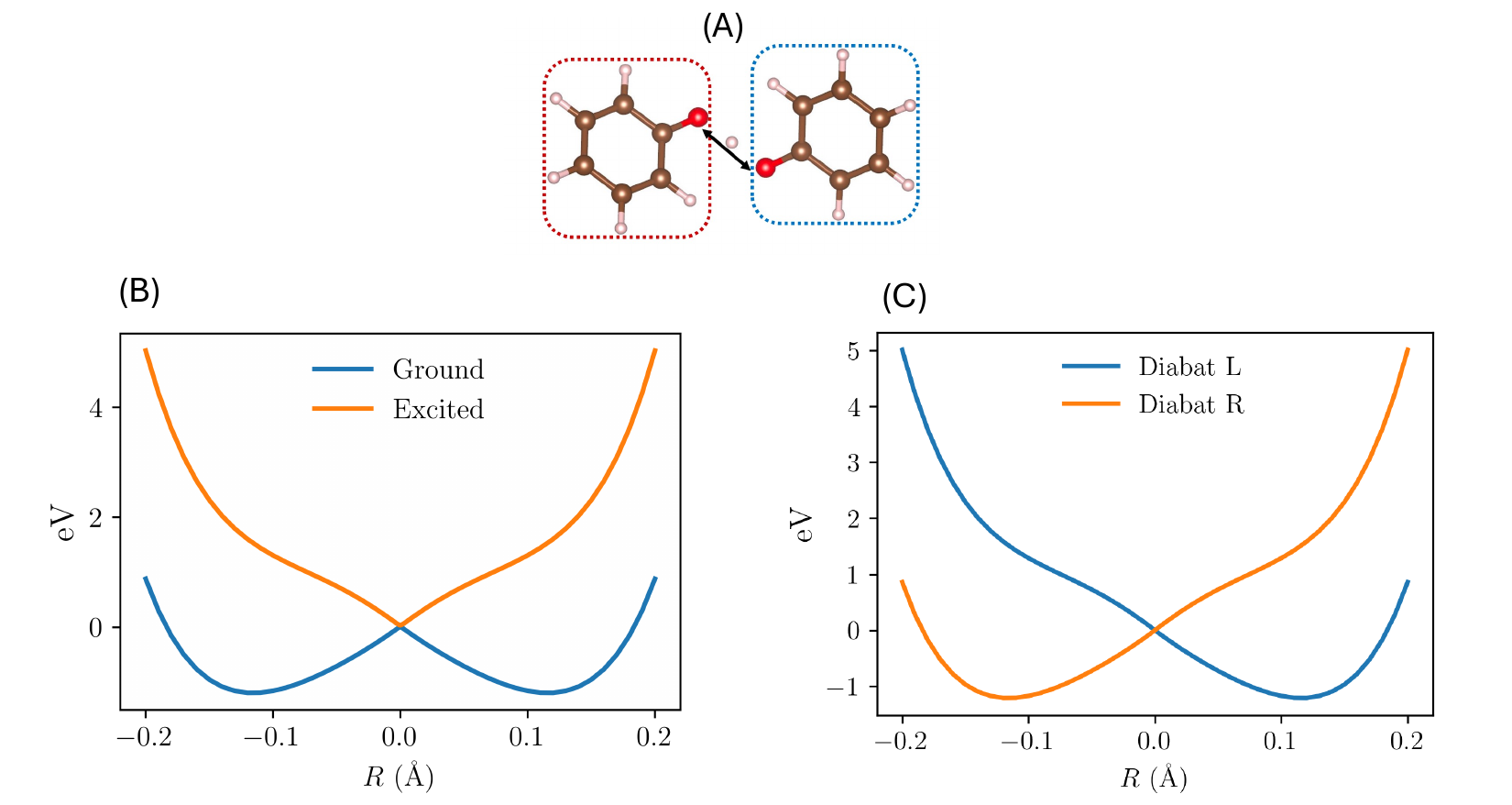}
    \caption{Potential energy curves for charge transfer along the reaction coordinate $R$ corresponding to a proton coupled electron transfer for a phenoxy-phenol (ph-ph) complex. (A) A visualization of the system whereby the bridging hydrogen is transfers between the two symmetrically located fragments. The internuclear distance between oxygen atoms is 2.459 \AA. The numerical value on the x-axis  represents the displacement of the hydrogen from the midpoint between the two oxygens. (B) Adiabatic surfaces.  (C) The  left (L) and right (R) diabatic surfaces. The charge transfer diabatic states are doubly degenerate (just like the adiabatic states).}
    \label{fig:pes_adiab_vs_diab}
\end{figure}

\begin{figure}[h!]
    \centering
    \includegraphics[width=0.6\textwidth]{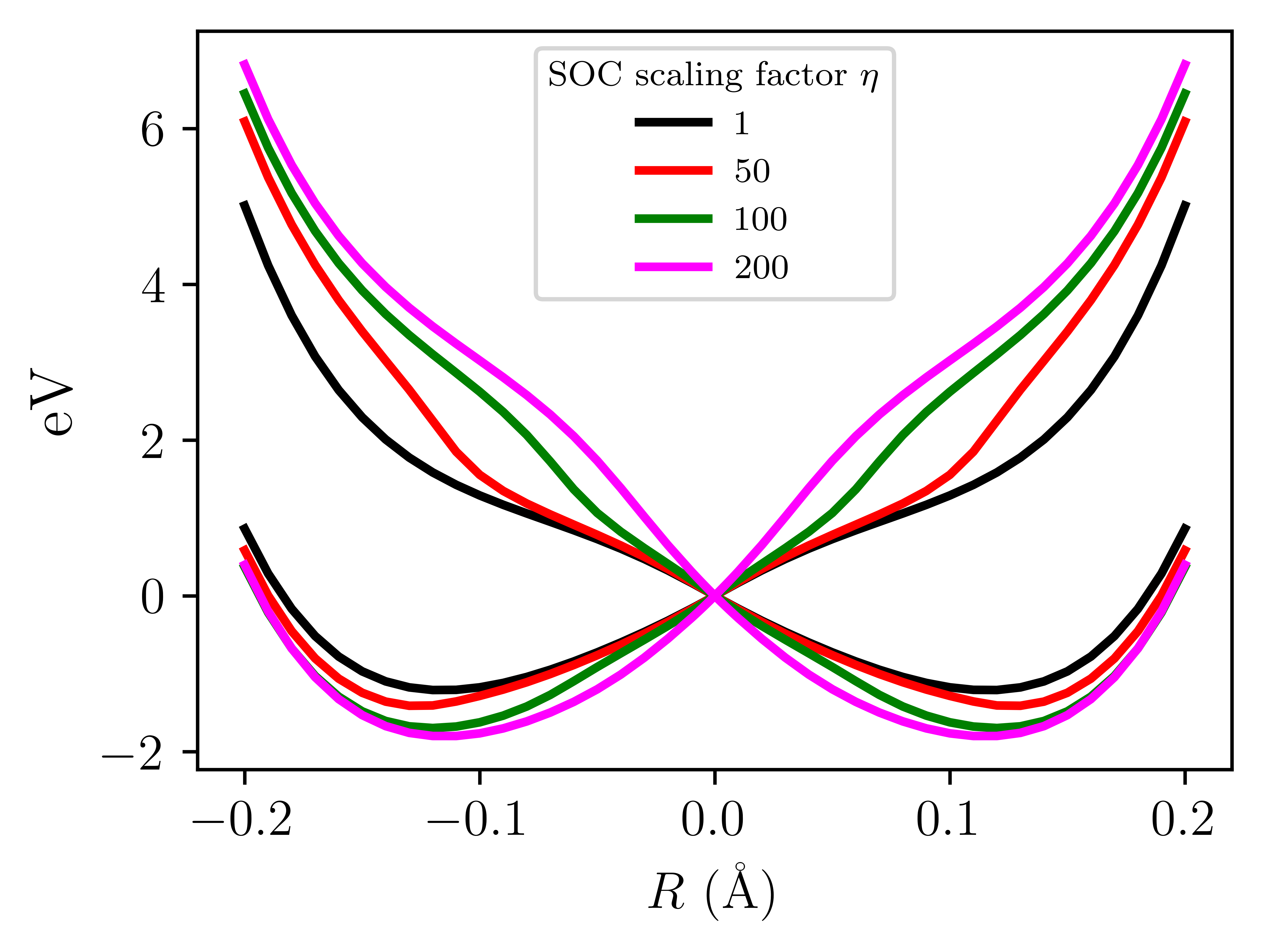}
    \caption{Diabatic PESs for different SOC values with scaling factor $\eta$ = [1, 50, 100, 200]. Details about  the form of spin-orbit coupling are in Paper~1 \cite{kumar2026chargetransferspini}. Note that increasing the SOC can drastically change the surfaces.
    \label{fig:pes_socX}}
\end{figure}

\begin{figure}[h!]
    \centering

    \begin{subfigure}{\textwidth}
        \centering
        \includegraphics[scale=0.55]{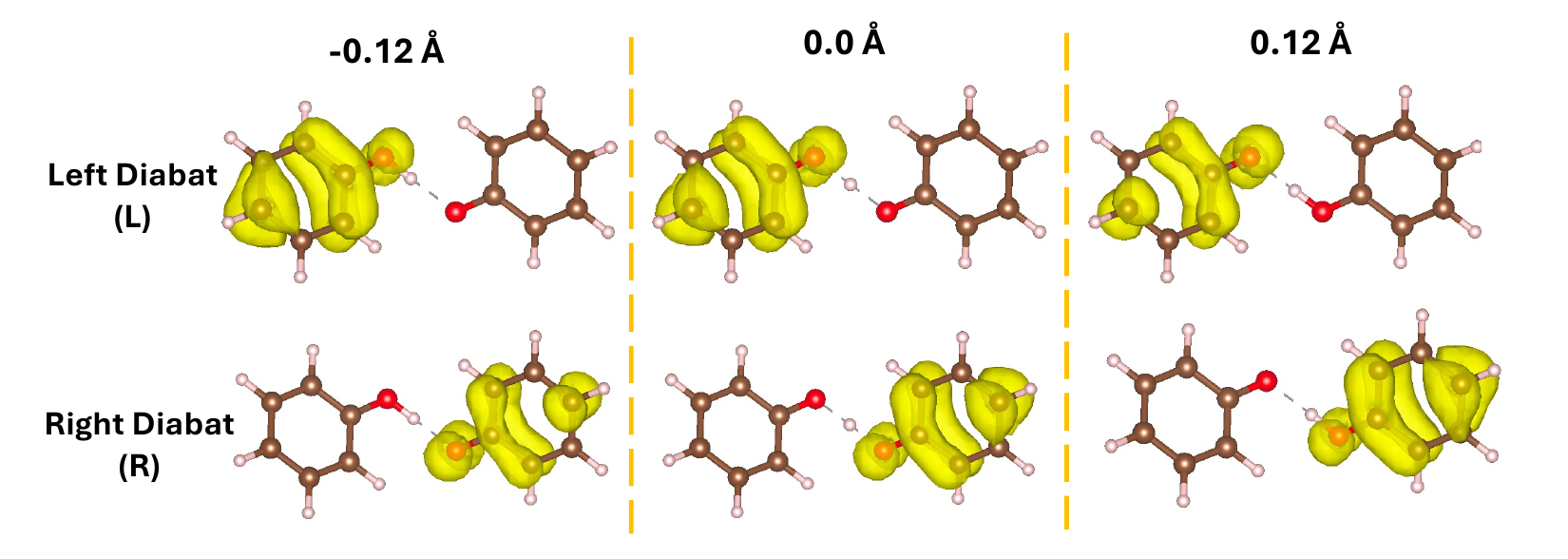}
        \caption{Active orbital electron density with SOC strength $\eta = 1$.}
        \label{fig:diabcube_soc1}
    \end{subfigure}

    \vspace{1em} 

    \begin{subfigure}{\textwidth}
        \centering
        \includegraphics[scale=0.55]{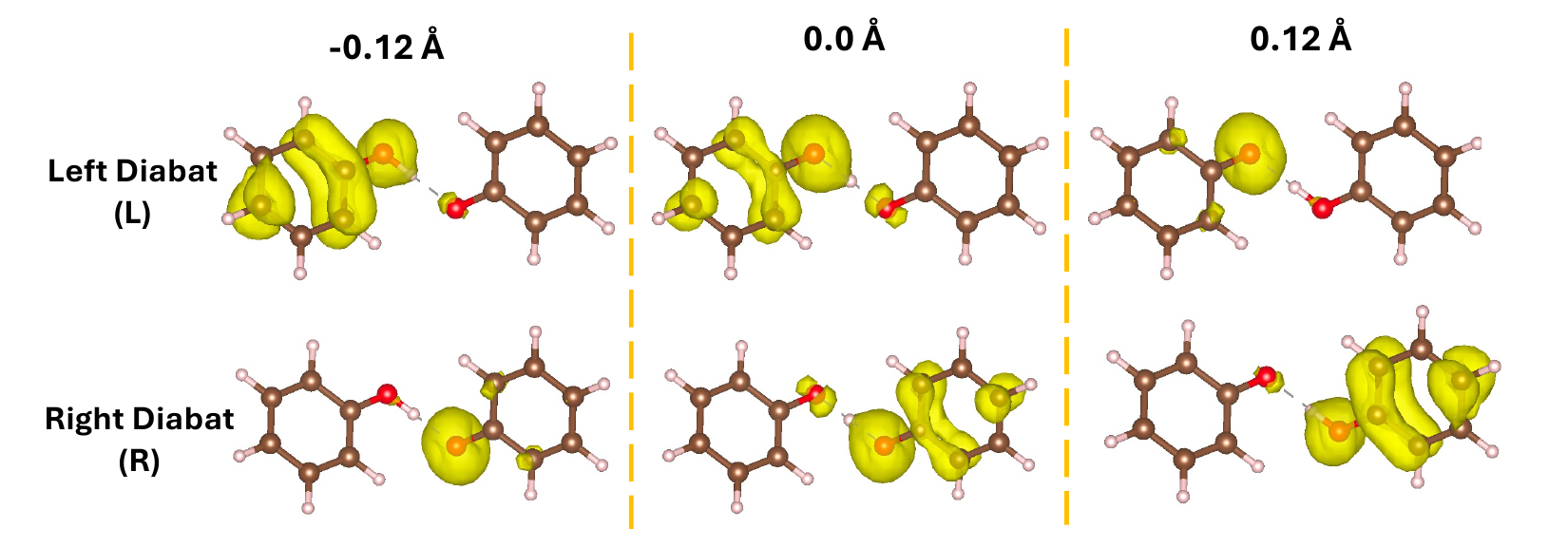}
        \caption{Active orbital electron density with SOC strength $\eta = 200$.}
        \label{fig:diabcube_soc200}
    \end{subfigure}

    \caption{A comparison of the active orbital electron densities for two charge transfer diabatic states at different SOC coupling strengths (A) $\eta = 1$ and (B) $\eta = 200$. Plotted at R = [-0.12 \AA, 0.0 \AA, 0.12 \AA] 
    }
    \label{fig:cube_soc_combined_diabat}
\end{figure}

As shown in Fig. \ref{fig:pes_adiab_vs_diab}, 
the diabatization procedure described above yields smooth, physically meaningful potential energy surfaces that exhibit well-defined crossings. 
To assess the robustness of the diabatization scheme, we applied the latter to the large range of spin–orbit coupling (SOC) strengths explored in Paper 1 \cite{kumar2026chargetransferspini}. As shown in Fig.~\ref{fig:pes_socX}, the resulting diabatic potential energy surfaces remain smooth and exhibit consistent crossing behavior over the entire range of SOC scaling factor $(\eta)$. Fig. \ref{fig:cube_soc_combined_diabat} compares the active-space electron densities of the two charge-transfer diabatic states at three representative geometries for $\eta$ values 1 and 200. For both SOC strengths, each diabat exhibits qualitatively distinct left- vs right-localized density. At strong SOC, the electron density varies more noticeably with geometry, indicating stronger mixing of orbital character along the reaction coordinate.

\newpage

\subsection{Spin expectation values and the spin quantization axis}
\label{sec:spin_expectation}

Next, for each diabatic state along the reaction coordinate, we compute the spin expectation vector 
\begin{align} \label{eq:spin_vec}
\mathbf{s}(\bm{R}) 
= \big(\langle \hat S_x\rangle,\langle \hat S_y\rangle,\langle \hat S_z\rangle \big),
\qquad
\langle \hat S_{\alpha}\rangle = \langle \xi(\bm{R}) | \hat S_{\alpha} | \xi(\bm{R})\rangle .
\end{align}

Note that the diabatic states are defined as unitary rotations within the projected space spanning two Kramers doublets. Note that, if the spin of one diabat points in the ${\bm s}$ direction, the spin of the other diabat points in the ${-\bm s}$  direction. Henceforth, we will refer to the 3D axis common to these two spin directions as a spin quantization axis.

\begin{figure}[ht]
    \centering
    \includegraphics[width=1
    \textwidth]{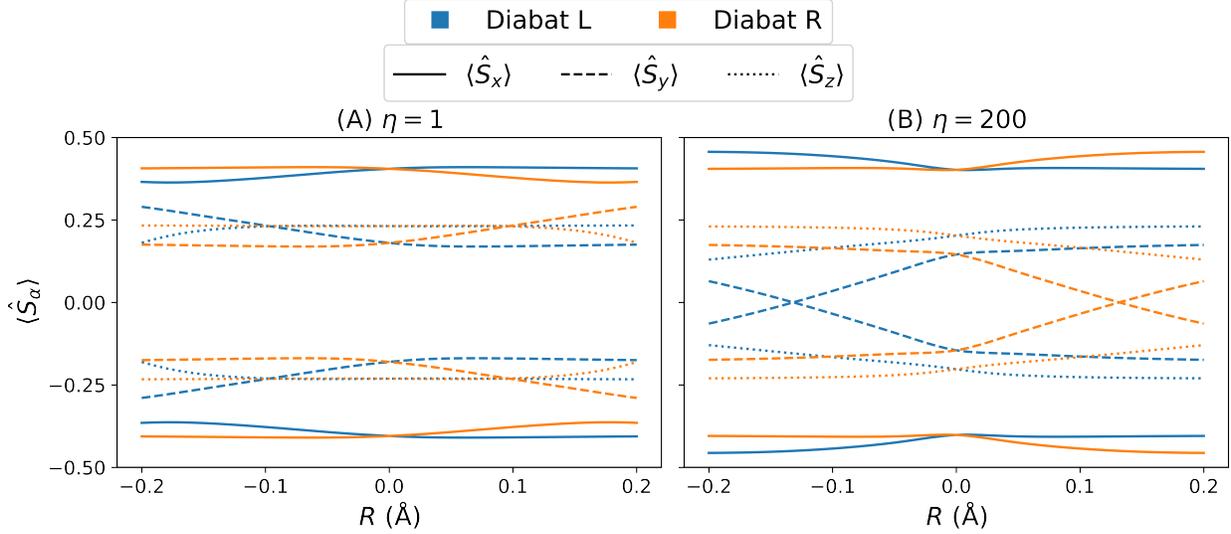}
    \caption{Spin expectation values $\langle \hat S_{\alpha}\rangle$ ($\alpha=x,y,z$) along the reaction coordinate for SOC strengths (A) $\eta=1$ and (B) $\eta=200$.
    Diabats L and R are shown in blue and orange, respectively. Solid, dashed, and dotted lines denote $\langle \hat S_x\rangle$, $\langle \hat S_y\rangle$, and $\langle \hat S_z\rangle$.
    For each diabat, the two time-reversal partners exhibit equal-and-opposite spin expectation values. See Fig. \ref{fig:mol_axis} for a definition of the $xyz$ axes.}
    \label{fig:sxyz_soc1_200} 
\end{figure}

\begin{figure}[h!]
    \centering

        \includegraphics[width=0.8\textwidth]{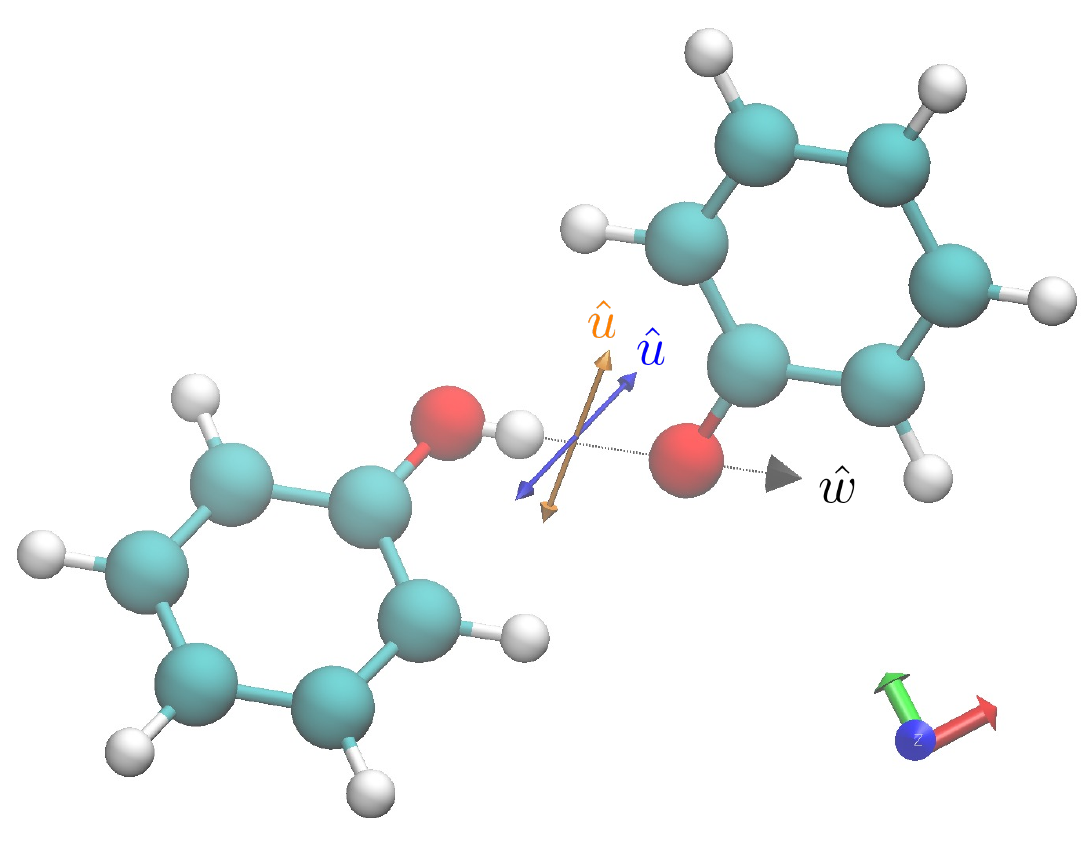}
        \caption{A visualization of the PCET process in phenoxy-phenol system, where hydrogen transfers between (red) oxygen fragments; the direction of hydrogen transfer defines the $\bm{\hat{w}}$ direction along the reaction coordinate. The blue and orange arrows (labelled $\bm{\hat{u}}$) are defined by spin expectation values in the body frame for the Left(L) and Right(R) diabats, respectively. See Eq. \ref{eq:proj_spin} for the relation between these vectors. The body frame directions are defined in bottom right with Z-axis coming out of the plane formed by red (X) and green (Y) arrows. The Cartesian coordinates are the same as used in Paper 1 \cite{kumar2026chargetransferspini}.}
        \label{fig:mol_axis}
\end{figure}

Figure \ref{fig:sxyz_soc1_200}
compares the Cartesian components of $\mathbf{s}(\bm{R})$ for weak ($\eta=1$) and strong ($\eta=200$) SOC strength, shown for the two localized diabats (L and R); for a definition of the $xyz$ directions, see Fig. \ref{fig:mol_axis}. In this diabatic basis, the spin expectation values vary smoothly with geometry, consistent with the intended charge-spin localization within the chosen four-state adiabatic basis. Because each Kramers doublet carries an intrinsic SU(2) gauge freedom, the Cartesian components $\langle \hat{S}_{\alpha} \rangle$ would normally exhibit spurious, rapidly varying fluctuations under different--yet physically equivalent--choices of gauge within the doublet. Our localized diabatic gauge fixes this rapidly varying gauge, leading to the smooth, slowly varying behavior.
Additionally, time-reversal symmetry is preserved in our diabatic representation, i.e., for each diabat the spin expectation values are equal in magnitude and opposite in sign for the time-reversal partners.

\begin{figure}[ht]
    \centering
        \includegraphics[width=1.0\textwidth]{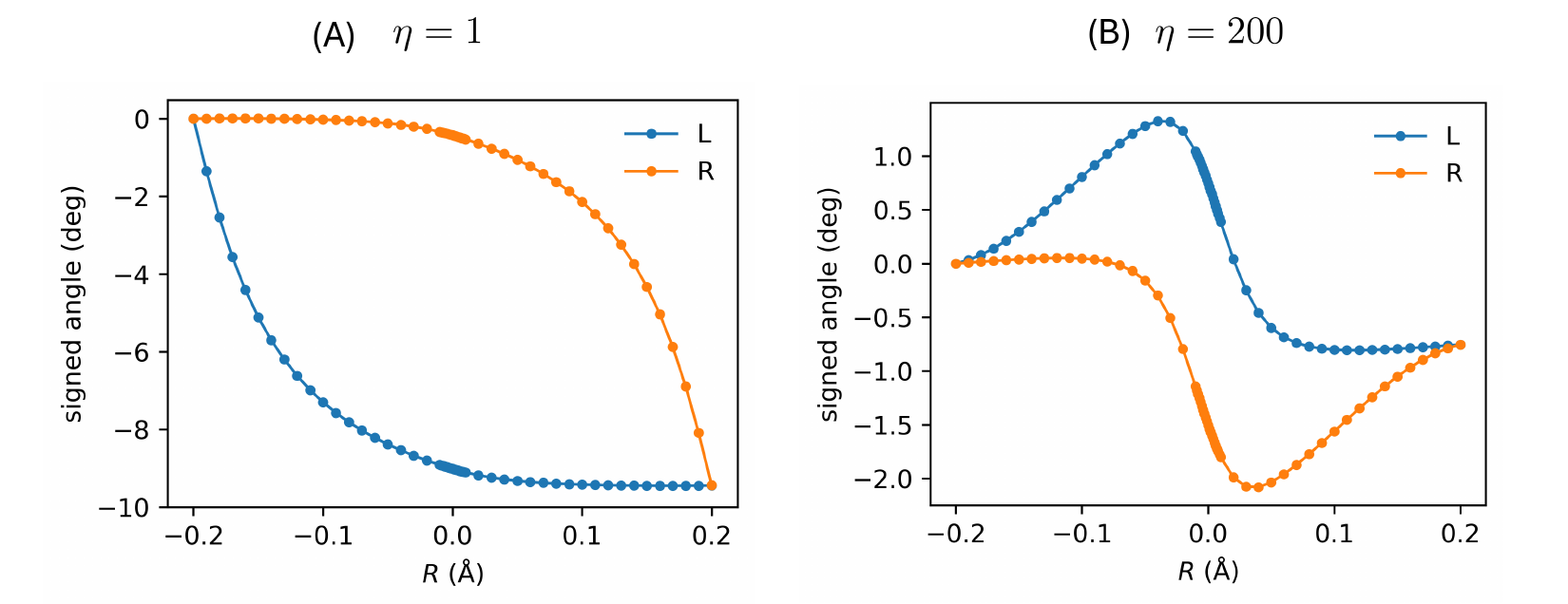}
        \caption{Signed angle $\theta(\bm{R})$ with reference spin vector chosen at $\bm{R}_{\mathrm{ref}}=-0.2 \textup{~\AA}$ geometry. Each point shows the signed rotation of the spin quantization axis  relative to $\bm{R}_{\mathrm{ref}}$, where the spin quantization axis is projected into the plane perpendicular to the reaction coordinate. (A) $\eta=1$ and (B) $\eta=200$.}
        \label{fig:signedangle_soc1_200}
\end{figure}

While Fig.~\ref{fig:sxyz_soc1_200}
reports the component-wise evolution of the spin
expectation vector in Eq. \ref{eq:spin_vec}, it is often more informative to quantify how the \emph{direction} of $\mathbf{s}(\bm{R})$ rotates along the
reaction coordinate.
Specifically, we let the position of bridging hydrogen vary  and we choose a viewing axis $\hat{\mathbf w}$ (here the O--O internuclear axis between two oxygen in Fig. \ref{fig:mol_axis} ) and remove
the component of $\mathbf{s}(\bm{R})$ parallel to $\hat{\mathbf w}$,
\begin{align}
\mathbf{s}_\perp(\bm{R}) &= \mathbf{s}(\bm{R})- \big(\mathbf{s}(\bm{R})\cdot \hat{\mathbf w}\big)\hat{\mathbf w}, \qquad
\hat{\mathbf u}(\bm{R}) = \frac{\mathbf{s}_\perp(\bm{R})}{\|\mathbf{s}_\perp(\bm{R})\|},
\label{eq:proj_spin}
\end{align}
so that $\hat{\mathbf u}(\bm{R})$ is a unit vector lying in the plane perpendicular to $\hat{\mathbf w}$.
We then compute the relative signed angle between $\hat{\mathbf u}(\bm{R}_{\rm ref})$ and $\hat{\mathbf u}(\bm{R})$ using a smooth
$\mathrm{arctan}$ convention (atan2 by C++),
\begin{align}
\theta(\bm{R}) = \mathrm{atan2}\!\left(
\hat{\mathbf w}\cdot\big(\hat{\mathbf u}(\bm{R}_{\rm ref})\times \hat{\mathbf u}(R)\big),\;
\hat{\mathbf u}(\bm{R}_{\rm ref})\cdot \hat{\mathbf u}(\bm{R})
\right),
\label{eq:signed_angle}
\end{align}
which returns a signed rotation relative to the reference geometry at $\bm{R}_{\rm ref}$.
The sign is fixed by the right-hand rule about $\hat{\mathbf w}$, allowing clockwise and
anticlockwise rotations to be distinguished. Figure \ref{fig:signedangle_soc1_200} uses the signed angle $\theta(\bm{R})$ to quantify, for both weak and strong SOC, how the spin quantization axis rotates as the molecule progresses along the reaction coordinate.

Overall, note that the diabatization procedure from Sec. \ref{sec:diabatization_cs} is able to reproduce a smooth and reasonably slowly changing spin axis as a function of nuclear position, offering us a reasonable starting frame for modeling spin effects during electron transfer (with minimal derivative couplings induced between different spin directions).

\section{Discussion}\label{sec:discussion}

\subsection{Spin as a good quantum number and time-reversal symmetry}

Although the presence of SOC implies that spin is no longer a good quantum number,  time-reversal symmetry (TRS) does still hold  in the absence of
external magnetic fields.  On the one hand, TRS arises because $[\hat H,\hat T]=0$ where $T$ is the anti-linear time-reversal operator.  For odd-electron (half-integer) systems, $\hat T^2=-1$ and TRS implies Kramers degeneracy.  On the other hand, spin is a good quantum number only for Hamiltonians that satisfy 
$[\hat H,\hat S_x]=[\hat H,\hat S_y]=[\hat H,\hat S_z]=[\hat H,\hat S^2]=0$, leading to
a global SU(2) spin-rotation invariance generated by
$\hat{S}$. Obviously, this latter symmetry is broken in the presence of SOC.

Within the realm of standard (variational) electronic structure calculations \cite{menon2008consequences, casanova2020spin, buvcinsky2015spin, kasper2020perspective}, one often faces a trade-off: enforcing the exact symmetries of the electronic Hamiltonian yields symmetry-adapted yet high energy solutions, whereas relaxing such constraints often gives a lower energy yet symmetry broken state with unphysical quantum numbers (e.g., spin contamination);
this tradeoff is termed  Löwdin’s ``symmetry dilemma'' \cite{lykos1963discussion}. Note that, for the method described in Paper~1 \cite{kumar2026chargetransferspini}, there is no spin contamination in the absence of SOC; deviations from $\ev{\hat{S}^2} = 3/4$ arise only when SOC is included and these deviations are then physically correct. 

\subsection[s square]{$\ev{\hat{S}^2}$ along a reaction coordinate}

With the background above in mind,  
it makes sense to calculate  the expectation value of $\hat{S^2}$ in order to learn about the actual effect of the SOC on the electronic structure.
Again, {\em without any SOC}, one should always recover $\ev{S^2} = 3/4$ for a doublet, and any deviation from this value is the result of spin contamination. 
To that end, we have analyzed $\ev{\hat{S}^2}$ values  across the reaction coordinate. As established in Ref.\citenum{cassam2015spin}, for a single Slater determinant based on Generalized Hartree-Fock (GHF) 
wavefunction, the expectation value of $\hat{S}^2$ is:
\begin{align}
\langle \psi_{i} | \hat{S}^2 | \psi_{i} \rangle 
&= \left( \frac{N_\alpha}{2} - \frac{N_\beta}{2} \right) 
   \left( \frac{N_\alpha}{2} - \frac{N_\beta}{2} + 1 \right) 
   + \frac{1}{4} \left( N_e - \sum_{i,j=1}^{N_e} 
   \big| \langle \psi_{i\alpha} | \psi_{j\alpha} \rangle 
     - \langle \psi_{i\beta} | \psi_{j\beta} \rangle \big|^2 \right) 
     \notag \\ &\quad 
     + \left( N_\beta - \sum_{i,j=1}^{N_e} 
   \langle \psi_{i\alpha} | \psi_{j\beta} \rangle 
   \langle \psi_{j\beta} | \psi_{i\alpha} \rangle \right) 
   + \left| \sum_{i=1}^{N_e} 
   \langle \psi_{i\beta} | \psi_{i\alpha} \rangle \right|^2 ,
\label{eq:s2_ghf}
\end{align}
where each GHF spin-orbital is a two-component spinor 
\begin{align}
\ket{\psi_{i}} \equiv 
\begin{pmatrix}
\ket{\psi_{i\alpha}} \\[4pt]
\ket{\psi_{i\beta}}
\end{pmatrix}   . 
\end{align}

Eq.~\ref{eq:s2_ghf} has been derived in detail in Ref.\citenum{cassam2015spin} and decomposes the GHF spin expectation value into four distinct contributions. In particular, the first term is analogous to Restricted Open-Shell Hartree-Fock (ROHF) expression where the “number of $\alpha$-spin electrons” (respectively “number of
$\beta$-spin electrons”) are defined as 
\begin{align}
    N_\alpha := \sum_{i=1}^{N_e} \langle \psi_{i\alpha} \vert \psi_{i\alpha} \rangle 
\qquad\text{(respectively,}\quad
N_\beta := \sum_{i=1}^{N_e} \langle \psi_{i\beta} \vert \psi_{i\beta} \rangle\text{)}    
\end{align}
where $N_e = N_{\alpha} + N_{\beta}$ is the total number of electrons.

\begin{figure}[ht]
    \centering
    \includegraphics[width=\textwidth]{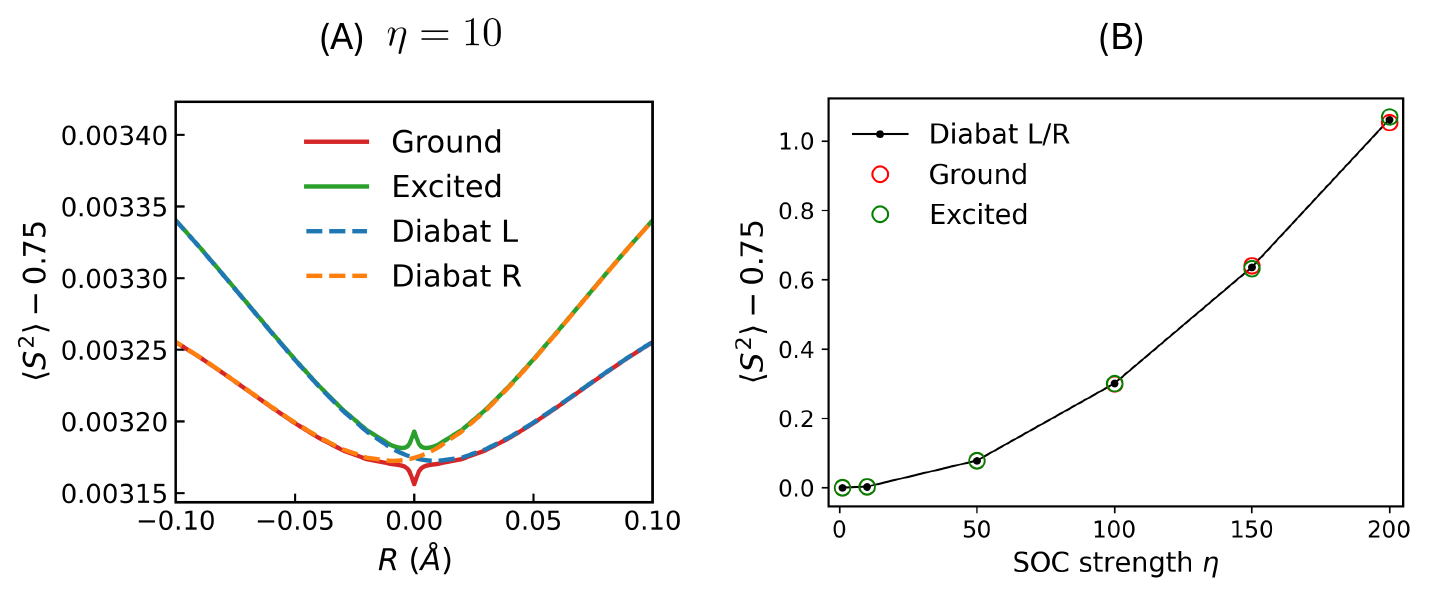}
    \caption{A plot of $\ev{\hat{S}^2}$ relative to  0.75.   (A) We plot $\ev{\hat{S}^2}$ for strength $\eta$ = 10 along the reaction coordinate. Here, we plot the values for both adiabatic and diabatic curves; mixing at $R=0$ does not change $\ev{\hat{S}^2}$ very much for this value of $\eta$. (B) We plot $\ev{\hat{S}^2}$ a function of SOC strength $\eta$ for diabatic states at the central geometry (R = 0.0 \AA)  where the diabats cross.  Note that scaling the SOC strength does not represent a physical reality of any kind; rather, the graphs  here are intended both  to assess the stability or our algorithm and to give some intuition as to what we might find more generally for large perturbations. Most importantly, in (B), we find that diabatization cannot restore spin as a good quantum number.
    }
    \label{fig:S2_cont_soc10}
\end{figure}

Results for  $\ev{\hat{S}^2}$  are plotted in Fig. \ref{fig:S2_cont_soc10}. We find two conclusions. First, note the scale on the y-axis of Fig. \ref{fig:S2_cont_soc10} (A) which plots the deviations before and after dipole diabatization (subsequent spin diabatization does not alter this value). Clearly, with minimal SOC ($\eta$ = 10), $\ev{\hat{S}^2}$  of the resulting ground state changes by only $0.05\% = 0.003/0.75$.  Thus, for the system studied here, we do not expect large SOC effects in general; larger effects should arise with more oxygen atoms and/or heavy elements.

Second and more importantly, when the states cross at  R = 0.0 \AA, we find only a very small change in $\ev{\hat{S}^2}$  between diabat and adiabat, with a relative change of $0.00002/0.00320 = 0.006\%.$  In Panel (B) of Fig. \ref{fig:S2_cont_soc10} , we plot the deviation at diabatic crossing for different SOC strength. We find that, although the total $\ev{\hat{S}^2}$  value strongly changes with SOC, the diabatic and adiabatic states largely agree as to their relevant $\ev{\hat{S}^2}$  values. In other words, if we use the procedure above to generate initial and final states for a spin-dependent fermi golden rule calculation, we should not expect the diabatization procedure itself will make spin a better quantum number; the entanglement of the spin and electronic degrees of freedom cannot be undone by a simple rotation operator of the relevant adiabats.

Lastly, for a general comparison, the $\ev{\hat{S}^2}$ value (at diabatic crossing) for a UHF calculation  without SOC is 1.475,  a significant deviation of more than 96 \%, arising from unphysical spin contamination. The magnitude of this number simply reminds the reader just how important it will be to use algorithms that exactly respect TRS when modeling spin dynamics in the future.

\section{Conclusion}\label{sec:conclusion}
We have developed and implemented a unified framework for constructing diabatic states that are simultaneously localized in both charge and spin for open-shell molecular systems subject to spin–orbit coupling. By combining dipole- and spin-based localization criteria within a sequence of complex-unitary Jacobi sweeps, the method achieves smooth potential energy surfaces that maintain Kramers degeneracy and preserve key symmetry properties.
Moreover, our diabatization framework provides a physical picture in which spin localization is understood as aligning the spin–quantization axis  smoothly and as steadily as possible along the reaction coordinate. 

The most crucial next question of this research direction is: what are the implications of this changing spin as far as electron transfer is concerned. In particular, there is a growing consensus that, in chiral environments, electron transfer can have a spin  preference, resulting in so-called ``chiral induced spin selectivity.''\cite{naaman2024can, eckvahl2023direct} To that end,   
one would very much like to explore how the ``chirality of the reaction path'' affects rotations in the spin quantization axis. 
And given that any changes in the spin quantization axis must originate from SOC, can we correlate such changes with orbital helicity \cite{bro2021quantification, garner2023fundamental} and what are the consequences for charge–transfer reactions? 

Finally, perhaps most interestingly, in order to more deeply probe CISS, one would like to explore the diabatic states described above in the context of an external magnetic field which may well lead to a spin preference for ET. In addition, recent explorations of phase space electronic structure theory\cite{bian2026phase, tao2025basis, bradbury2025symmetry} have indicated that one can gain additional accuracy in modeling nuclear-electronic correlation by using a zeroth order electronic Hamiltonian that depends on both nuclear position $R$ and nuclear momentum $P$, and thus includes an internal magnetic field. The diabatic state basis above would appear to be an obvious starting point for more deeply exploring CISS within such a phase space framework.

\section{Appendix}\label{sec:si}

\subsection{Jacobi sweeps}\label{sec:si_jacobi_sweep}

Given \(N(N>1)\) Hermitian operators \(\bm{O}^{(k)}\) and an orthonormal basis \(\{\psi_i \} \), we want to find the maximum of the functional
	\begin{align}\label{eq:JacobiSweep:func}
	f(U) = \sum_{i} \sum_{k}\left( \left < {\xi_i} | {\bm{O}^{(k)}} |{\xi_i} \right>\right)^2
	\end{align}
    
	where
	\begin{align}\label{}
	\ket{\xi_j} = \sum_{i} \ket{\psi_i} U_{ij}
	\end{align}
	and $\bm{U}$ is a unitary matrix. In this work, $\bm{O}^{(k)}$ represents the set $\{ \bm{\mu}_x, \bm{\mu}_y, \bm{\mu}_z \}$ and $\{ \bm{S}_x, \bm{S}_y, \bm{S}_z \}$ for dipole and spin diabatization, respectively.
    
    Numerically this optimization task can be done by successive Jacobi rotations between a pair of states using matrix $\bm{U}$. An iteration over all pairs is called a Jacobi sweep. We iterate until convergence of the maximization function in Eq. \ref{eq:JacobiSweep:func}. 
	
	Let us now focus on a single Jacobi rotation step. Consider a 2-by-2 rotation within the subspace formed by \( \{\xi_i, \xi_j \} \)
	\begin{align}
	\begin{bmatrix}
	\xi_i' & \xi_j'
	\end{bmatrix} = \begin{bmatrix}
	\xi_i & \xi_j
	\end{bmatrix} \begin{bmatrix}
	\cos\gamma & -\sin\gamma e^{-i\phi}\\
	\sin\gamma e^{i\phi} & \cos\gamma
	\end{bmatrix}
	\end{align}\label{eq:update_state}
For each operator \(O^{(k)}\), the change of Eq. \ref{eq:JacobiSweep:func} from this rotation is
    \begin{align}\label{}
	\Delta f^{(k)} ={}& \left[ O_{ii}^{(k)} \cos^2\gamma  + O_{jj}^{(k)}\sin^2\gamma + \left(O_{ij}^{(k)}e^{i\phi}+O_{ji}^{(k)} e^{-i\phi}\right)\cos\gamma\sin\gamma\right]^2 \nonumber \\
	&+ \left[ O^{(k)}_{ii}\sin^2\gamma + O^{(k)}_{jj}\cos^2\gamma - \left(O^{(k)}_{ij}e^{i\phi} + O^{(k)}_{ji}e^{-i\phi}\right) \cos\gamma\sin\gamma \right]^2 - (O^{(k)}_{ii})^2 - (O^{(k)}_{jj})^2 \nonumber \\
    \end{align}
Defining 
	\begin{align}\label{}
	A & \equiv \sum_{k} \left[ \left(\frac{O_{ij}^{(k)}e^{i\phi}+O_{ji}^{(k)}e^{-i\phi}}{2} \right)^2-\left(\frac{O^{(k)}_{ii}-O^{(k)}_{jj}}{2}\right)^2\right] \\
	B & \equiv \sum_{k} \frac{O_{ij}^{(k)}e^{i\phi}+O_{ji}^{(k)}e^{-i\phi}}{2} (O^{(k)}_{ii}-O^{(k)}_{jj}) 
    \end{align}
Then total change of Eq. \ref{eq:JacobiSweep:func} from this rotation is
	\begin{align}\label{eq: cossin}
	\Delta f = \sum_{k} \Delta f^{(k)} =  A - A\cos 4\gamma  + B\sin 4\gamma  = A + \sqrt{A^2+B^2} \cos 4(\gamma-\alpha)
	\end{align}
	where
	\begin{align}\label{}
	\cos 4\alpha &= -A/\sqrt{A^2+B^2}\\
	\sin 4\alpha &= B / \sqrt{A^2+B^2}
	\end{align}
	for the real case (where \(\phi\equiv 0\)) the maximum is readily obtained by choosing \(\gamma=\alpha+n\pi/2 \) where \(n\) is an arbitrary integer. For the complex case, one needs to first maximize \(A+\sqrt{A^2+B^2}\) with respect to \(\phi\). 

    	To maximize \(A +\sqrt{A^2+B^2}\), we first rewrite \(A\) and \(B\):
	\begin{align}\label{}
	A &= \frac{1}{4}\sum_{k}\left[  (O_{ij}^{(k)})^2 e^{2i\phi} + (O_{ji}^{(k)})^2e^{-2i\phi} + 2O_{ij}^{(k)}O_{ji}^{(k)}  -  (O_{ii}^{(k)} - O_{jj}^{(k)})^2 \right]\nonumber \\
	&= \frac{1}{4} \left[e^{2i\phi} \sum_{k} (O_{ij}^{(k)})^2 + \text{c.c.} \right] + \sum_{k} \left[\frac{1}{2}O_{ij}^{(k)}O_{ji}^{(k)} - \frac{1}{4} (O_{ii}^{(k)}-O_{jj}^{(k)})^2 \right] \nonumber  \\
	&\equiv u e^{2i\phi} + u^* e^{-2i\phi} + A_0 \label{eq:JacobiSweep:A}\\ 
	B&= \frac{1}{2} \left[e^{i\phi} \sum_{k} O_{ij}^{(k)} (O_{ii}^{(k)}-O_{jj}^{(k)}) + \text{c.c.}  \right] \nonumber \\
	&\equiv v e^{i\phi} + v^* e^{-i\phi} \label{eq:JacobiSweep:B}
	\end{align}
	where
	\begin{align}\label{}
	u&\equiv \frac{1}{4} \sum_{k} (O_{ij}^{(k)})^2\\
	v&\equiv \frac{1}{2}  \sum_{k} O_{ij}^{(k)} (O_{ii}^{(k)}-O_{jj}^{(k)}) \\
	A_0&\equiv \sum_{k} \left[\frac{1}{2}O_{ij}^{(k)}O_{ji}^{(k)} - \frac{1}{4} (O_{ii}^{(k)}-O_{jj}^{(k)})^2 \right] 
	\end{align}
	\(u\) and \(v\) in general are complex numbers, and \(A_0, A\) and \(B\) are real.
	The \(\phi\) that maximizes \(A +\sqrt{A^2+B^2}\) can be solved by the following equation:
	\begin{align}\label{eq:JacobiSweep:deriv}
	&\dv{\phi} \left(A +\sqrt{A^2+B^2}\right) = 0 \nonumber \\
	\implies & -\sqrt{A^2+B^2} \dv{A}{\phi} = A\dv{A}{\phi} + B \dv{B}{\phi} \nonumber \\
	\implies & B^2 \left[\left(\dv{A}{\phi}\right)^2- \left(\dv{B}{\phi}\right)^2\right] = 2AB\dv{A}{\phi}\dv{B}{\phi} \nonumber \\
	\implies & B \left[\left(\dv{A}{\phi}\right)^2- \left(\dv{B}{\phi}\right)^2\right] = 2A\dv{A}{\phi}\dv{B}{\phi}
	\end{align}
	where we've assumed \(B\neq 0\).
	
	Before we plug Eqs. \ref{eq:JacobiSweep:A} and \ref{eq:JacobiSweep:B} into Eq. \ref{eq:JacobiSweep:deriv}, it's helpful to define a few notations to make the expression looks simpler. Let
	\begin{align}\label{}
		x &\equiv u e^{2i\phi} \label{eq:JacobiSweep:x}\\
		y &\equiv v e^{i\phi} \label{eq:JacobiSweep:y}
	\end{align}
	then
	\begin{align}\label{}
	A &= x + x^* + A_0 \label{eq:JacobiSweep:A:x}\\
	B&= y + y^*\\
	\dv{A}{\phi} &= 2i\left(ue^{2i\phi} - u^* e^{-2i\phi}\right) = 2i(x-x^*)\\
	\dv{B}{\phi} &= i\left(ve^{i\phi}-v^*e^{-i\phi}\right) = i(y-y^*)\label{eq:JacobiSweep:dB:y}
	\end{align}
	
    Plugging Eqs. \ref{eq:JacobiSweep:A:x}-\ref{eq:JacobiSweep:dB:y} into Eq. \ref{eq:JacobiSweep:deriv}, and if we rewrite \(x\) and \(y\)  in terms of \(u,v\) and \(e^{i\phi}\), we  obtain an equation for \(e^{i\phi}\),
\begin{align}\label{eq:JacobiSweep:eq}
		\alpha e^{3i\phi} + \beta e^{i\phi} + \beta^* e^{-i\phi} + \alpha^* e^{-3i\phi} = 0
	\end{align}
    where
	\begin{align}\label{}
		\alpha &\equiv  v^3 + 4A_0 uv -8u^2 v^* \\
		\beta &\equiv  -v^2 v^*-4A_0uv^* + 8uu^*v 
	\end{align}

Note that, in terms of the power of \(e^{i\phi}\), this equation contains only \(e^{3i\phi}, e^{i\phi}, e^{-i\phi}\) and \(e^{-3i\phi}\). Moreover, the coefficient of \(e^{-3i\phi}\) is the complex conjugate of the coefficient of \(e^{3i\phi}\); so does the relation between \(e^{i\phi}\) and \(e^{-i\phi}\). One way to look at Eq. \ref{eq:JacobiSweep:eq} is to combine the complex conjugated terms:
	\begin{align}\label{}
		&\Re{\alpha e^{3i\phi}} + \Re{\beta e^{i\phi}} = 0 \nonumber \\
		\implies & \abs{\alpha} \cos(3\phi+\phi_{\alpha}) + \abs{\beta} \cos(\phi+\phi_\beta) = 0 \nonumber \\
		\implies & \frac{\cos(3\phi+\phi_{\alpha})}{\cos(\phi+\phi_\beta)} = -\abs{\frac{\beta}{\alpha}}
	\end{align}

	
	Another way to look at Eq. \ref{eq:JacobiSweep:eq} is to convert it to a cubic equation
	\begin{align}\label{eq:JacobiSweep:cubic}
	\alpha \omega^3 + \beta \omega^2 + \beta^* \omega + \alpha^*=0
	\end{align}
	where
	\begin{align}\label{}
	\omega &\equiv e^{2i\phi}
	\end{align}

Finally, for each Jacobi rotation, we plug in all three roots back in Eq. \ref{eq:JacobiSweep:func} and pick the one which gives maximum value of the optimizing functional.

\section{Acknowledgments}
We thank Brian Conrad and Keith Conrad for helpful discussions regarding Eq. \ref{eq:JacobiSweep:eq}.
This work was supported by the National Science Foundation
Grant CHE-2422858.

\bibliography{cite_all}

\newpage
\begin{tocentry}
  \centering
  \includegraphics[
    width=\linewidth,
    height=1.75in,
    keepaspectratio
  ]{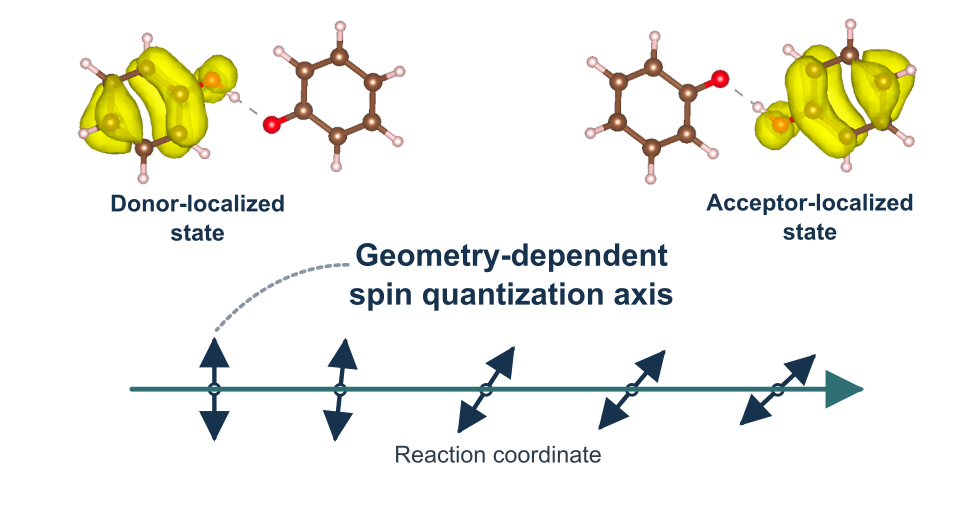}
\end{tocentry}
\end{document}